\documentclass[acmtog]{acmart} %

\usepackage{booktabs} %
\usepackage{xspace}
\usepackage{amsmath}
\usepackage{amsfonts}
\usepackage{cleveref}
\usepackage{microtype}
\usepackage{enumitem}
\setlist{topsep=2pt, leftmargin=*}

\newcommand{\revision}[1]{\textcolor{blue}{#1}}

\setcopyright{cc}
\setcctype{by}
\acmJournal{TOG}
\acmYear{2025} \acmVolume{44} \acmNumber{4} \acmArticle{xx} \acmMonth{8} \acmPrice{}\acmDOI{10.1145/3730926}

\graphicspath{ {figures/png_figs/} }

\newcommand{\name}{MonetGPT\xspace}
\newcommand{\llm}{$\mathcal{M}$\xspace}
\newcommand{\source}{$I_S$\xspace}
\newcommand{\edited}{$I_E$\xspace}
\newcommand{\expert}{$I_X$\xspace}
\newcommand{\adjust}{$A$\xspace}
\newcommand{\adjustvalue}{$V$\xspace}

\newcommand{\plan}{$\mathcal{P}$\xspace}
\newcommand{\library}{$\mathcal{L}$\xspace}
\newcommand{\reason}{$R$\xspace}
\newcommand{\operation}{$\mathcal{O}$\xspace}

\begin{document}

\title{\name: Solving Puzzles Enhances MLLMs' Image Retouching Skills}

\author{Niladri Shekhar Dutt}
\email{niladri.dutt.22@ucl.ac.uk}
\affiliation{%
  \institution{University College London}
  \country{UK}
}

\author{Duygu Ceylan}
\email{ceylan@adobe.com}
\affiliation{%
  \institution{Adobe Research}
  \country{UK}
}

\author{Niloy J. Mitra}
\email{n.mitra@ucl.ac.uk}
\affiliation{%
  \institution{University College London, Adobe Research}
  \country{UK}
}

\begin{abstract}
Retouching is an essential task in post-manipulation of raw photographs. %
Generative editing, guided by text or strokes, provides a new tool accessible to users but can easily change the identity of the original objects in unacceptable and unpredictable ways. In contrast, although traditional procedural edits, as commonly supported by photoediting tools (e.g., Gimp, Lightroom), are conservative, they are still preferred by professionals. Unfortunately, professional quality retouching involves many individual procedural editing operations that is challenging to plan for most novices. 
In this paper, we ask if a multimodal large language model~(MLLM) can be \textit{taught} to critique raw photographs, suggest suitable remedies, and finally realize them with a given set of pre-authored procedural image operations.  
We demonstrate that MLLMs can be first made aware of the underlying image processing operations, by training them to solve specially-designed visual puzzles. 
Subsequently, such an operation-aware MLLM can both plan and propose edit sequences. To facilitate training, given a set of expert-edited photos, we synthesize a reasoning dataset by procedurally manipulating the expert edits and then grounding a pretrained LLM on the visual adjustments, to synthesize reasoning for finetuning. 
The proposed retouching operations are, by construction, understandable by the users, preserve object details and resolution, and can be optionally overridden. We evaluate our setup on a variety of test examples and show advantages, in terms of explainability and identity preservation, over existing generative and other procedural alternatives. \textit{Code, data, models, and supplementary results can be found via our project website at \href{https://monetgpt.github.io/}{\revision{https://monetgpt.github.io}}. }

\end{abstract}


\begin{teaserfigure}
\includegraphics[width=\textwidth]{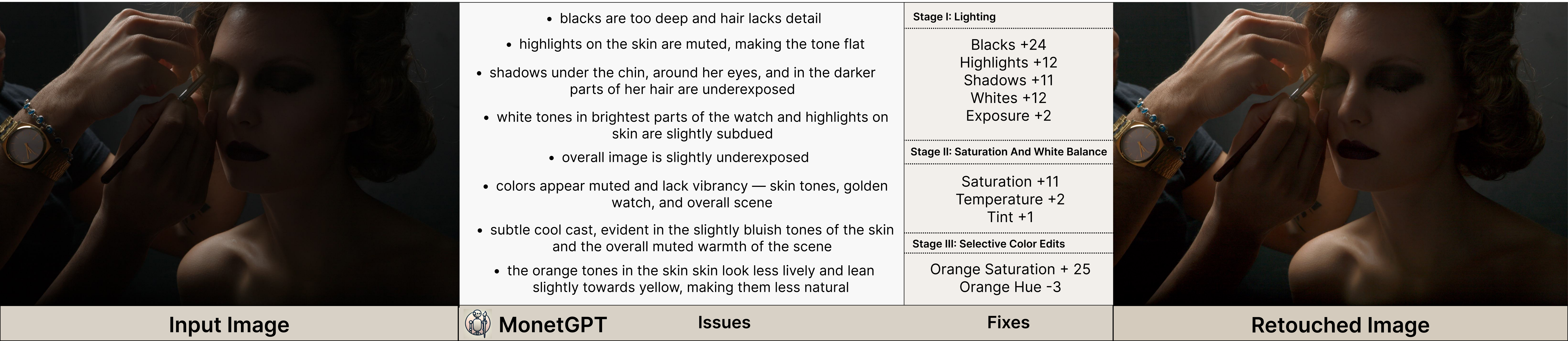}
  \caption{
  We present \name, an image operation-aware multimodal large language model~(MLLM), that provides automatic suggestions for image retouching. Given a photograph~(left), \name analyzes it to identify a set of issues and possible adjustments to fix them. The solution steps are then translated to a set of procedural operations, along with respective parameter settings, drawing from a given library of operations, which occurs in three stages. (Visual puzzles, on which we train the MLLM, are not shown here.)}
  \label{fig:teaser}
\end{teaserfigure}

\maketitle
\section{Introduction}
We regularly retouch captured images to improve their presentation. For example, users adjust contrast and brightness, manipulate exposure, or correct color profiles. Such adjustments, often consisting of a series of procedural operations, are preferred by professional users as the operations are non-destructive, and can be applied at different resolutions. Further, such edits are interpretable, supported by many established image manipulation tools, and, unlike generative editing counterparts, better preserve the identity of the source content.

Unfortunately, effectively using procedural edits is difficult and beyond the reach of most novices. There are two main challenges. 
First, users should learn how to apply the individual operations using the tool -- referred to as the \textit{command knowledge}. Second, they have to plan, based on the source image, which set of operations to use and propose appropriate parameter values for the chosen operations -- referred to as the \textit{strategic knowledge}. While the first one can be lowered with practice on a given toolset (e.g., Gimp), the second one is often hard to overcome as planning, using a library of operations, is open-ended and inherently more difficult. 

In a breakthrough effort, the Exposure framework~\cite{hu2018exposure} demonstrated that it is possible to learn, using a reinforcement learning setup, sequences of procedural edits directly from artist retouching examples. However, the effectiveness of such an approach is limited by the paucity of expert edits available for training. In this paper, we ask if it is instead possible to start from the knowledge embedded in frontier models, trained on significantly larger and diverse datasets, and adapt them to our specialized retouching task with limited data from expert artists.

We present \textit{\name} for procedural image retouching. \name introduces an effective fine-tuning strategy to adapt multimodal large language models~(MLLMs), even on a limited retouching dataset~\cite{ppr10k,Bychkovsky2011}. Once fine-tuned, the MLLM can identify problems in a source image, plan a sequence of fixes to improve the image, and finally translate the fixes using a given library of procedural edits. For example, as in \Cref{fig:teaser}, \name suggests an edit sequence to retouch the input raw photograph. In addition to proposing a sequence of edit operations with associated parameters, our method also provides \textit{explanation} in the form of what issue each of the proposed adjustments attempts to solve. Further, an user can ignore or override any of the proposed changes and run the rest of the procedure (e.g., experts sometimes violate photographic guidance to highlight an aspect/subject of the image).

\begin{figure}[t!]
    \centering
\includegraphics[width=\columnwidth]{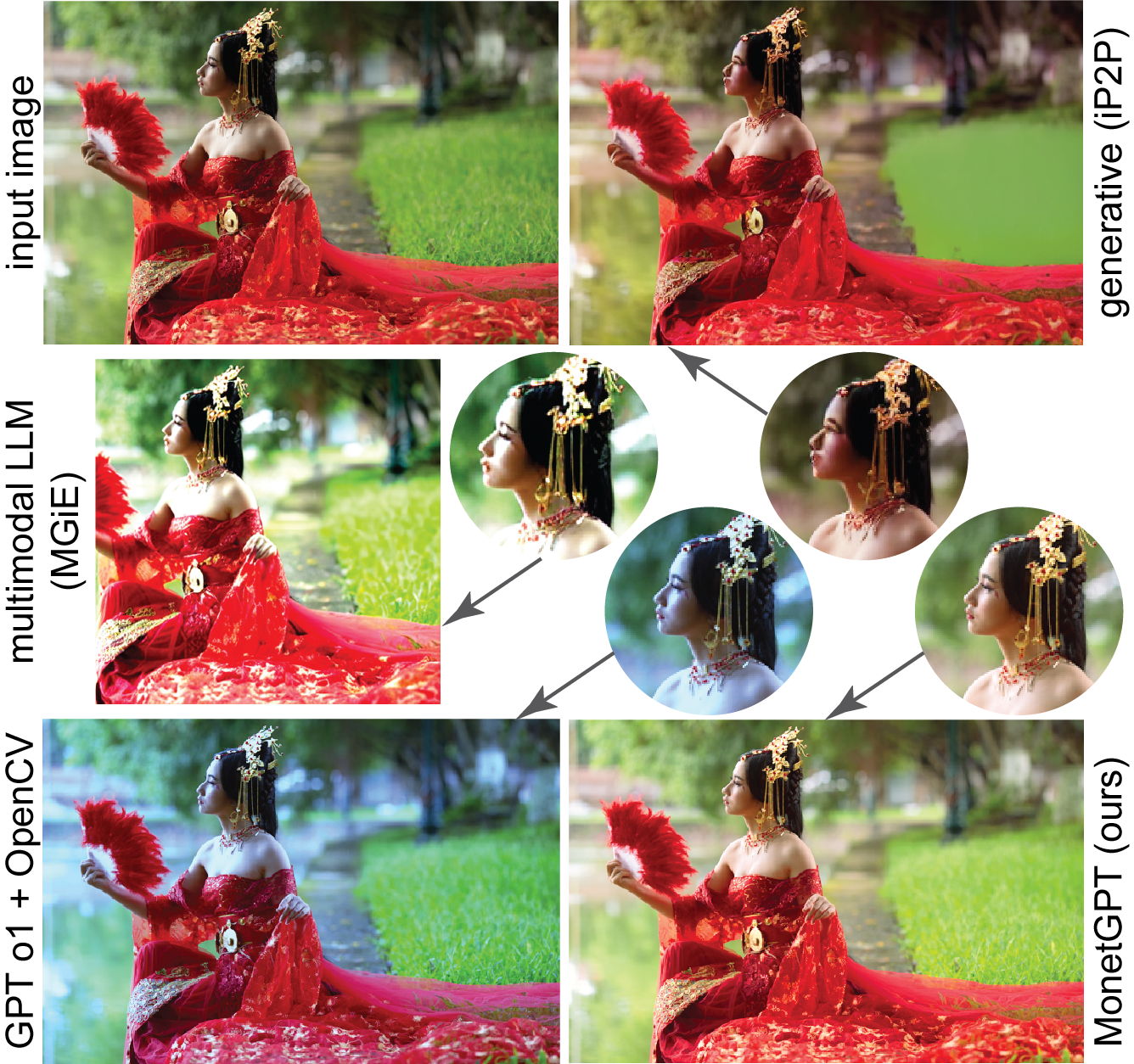}
    \caption{
    Generative tools, for example instructPix2Pix~\cite{brooks2022instructpix2pix} or MGIE~\cite{mgie24}, produce impressive image enhancements but can result in identity loss (e.g., faces, hands, objects) and are harder to override by users. 
    Procedural approaches are more controllable as they restrict operations to a given set of user-prescribed operation library, and can be overridden or applied in parts. 
    Current MLLMs (bottom-left: e.g., GPT o1 applied using a library of operations, presented as doc-strings), do not have a good internal model of image operations, and perform worse than our operation-aware variant~(bottom-right). See \Cref{sec:evaluation} for evaluation. 
    }
    \label{fig:motiovationGenerative}
\end{figure}

We found current MLLMs to be versatile but not quite powerful in proposing meaningful retouchings on raw photographs (see \Cref{fig:motiovationGenerative}). Directly fine-tuning such an MLLM on artists' edit sequences only partially improves the results (see \Cref{sec:evaluation}). We attribute this to MLLM's lack of understanding of what each of the image editing operations entails. As humans, we build mental models of these operations based on our experience (e.g., what does increasing the \texttt{brightness} slider do to images?). We mimic the same skill learning~\cite{skillLearn:20} for MLLMs -- to this end, we design specific visual puzzles, involving the given imaging operations, and train an MLLM on solving these puzzles. As a result the MLLM becomes operation-aware and is then capable of planning high-quality image retouching sequences. We demonstrate how to do so in an unpaired training setup with limited amount of data from artists.

We evaluated \name on a diverse set of input images, comparing its retouching results against generative editing methods (LLM-based editing, InstructPix2Pix~\cite{brooks2022instructpix2pix}, and MGIE~\cite{mgie24}), procedural editing techniques from prior work (Exposure~\cite{hu2018exposure}), custom-designed variations of MLLMs integrated with procedural pipelines (Gemini 2.0 with chain-of-thought, fine-tuned variants), and commercial software (GooglePhoto AutoEnhance). We evaluated the resultant edits 
extensively through qualitative analysis by expert reviewers and novice users. Please refer to supplemental for more evaluation. In summary, we 
\begin{enumerate}[label=(\roman*)]
\item propose the first framework for MLLM-guided interpretable procedural image retouching, enabling non-destructive editing on high-resolution 16-bit images.,
\item train MLLMs on strategically-designed \textit{visual puzzles} to make them operation-and aesthetics-aware, and use to plan our edit sequences with associated parameter estimation;
\item demonstrate, via extensive evaluation and comparisons, the feasibility and benefits of operation-aware MLLM-guided procedural editing workflows over chain-of-thought reasoning MLLMs and generative alternatives. 
\end{enumerate}

Refer to our webpage at \href{https://monetgpt.github.io/}{\revision{monetgpt.github.io}} for code and supplementary results.

\section{Related Work}

\subsection{Image Retouching}
Image retouching is an essential and commonly used workflow for post-processing raw images. Many commercial image editing software and web-based tools provide a vast number of filters that can be used for image enhancement. Given the difficulty in choosing which filters to apply in which settings, there has been a lot of research to automate parts of this workflow. A popular thread of research focuses on using pairs of input and edited images to predict the parameters of individual filters, such as global tone adjustment or color adjustment. Earlier work~\cite{Bychkovsky2011,Yan2014} used machine learning methods such as Gaussian process regression or support vector machines to tackle this task. Later, such methods were replaced with deep neural networks as surrogate functions for various image processing operations~\cite{Yan:adjust:16,Chen_2017_ICCV,Liu2023}. More recently, approaches that aim to perform image enhancement directly by predicting residual image layers~\cite{kim2020pienet} or per-pixel color and channel intensity transforms~\cite{Kim:2020} have been developed. Li et al.~\shortcite{Li_2023_CVPR} strike a balance between global and per-pixel editing by using a set of piecewise linear curves to retouch different spatial regions of an input image. However, in such approaches, it is not possible to further edit or control the results since the edits are not linked to specific image processing operations.

Closer to our problem setup, several works have looked into how to best select the type and parameters of a predefined set of operations to enhance image quality and aesthetics. Notably, Exposure~\cite{hu2018exposure} presents an RL-based framework where image retouching is cast as a planning problem: a discriminator that classifies an image as retouched or not is used to provide the reward function for the RL agent. In a similar setup, Shi et al.~\shortcite{Shi_2021_CVPR} propose to edit an image given a text prompt by generating a sequence of image editing operations and the corresponding parameters. The operations are chosen from a predefined stack of differentiable filters and a sequence modeler (i.e., LSTM decoder) is used to guide the planning.  Fischer et al.~\shortcite{fischer2020nicer} also propose a framework to optimize the parameters of a set of differentiable neural image filters using a neural image assessor to evaluate the image quality. Instead of training a planning algorithm from scratch, we explore the power of pre-trained multimodal large language models to aid learning from a limited set of expert edits.

\subsection{Generative Edits}
Over the past few years, there has been a transformational break-through in conditional and unconditional image generation, first with the use of GANs~\cite{gan2014} and most recently diffusion-based image generators~\cite{rombach2021highresolution}. Specifically, with the success of text-to-image generators, many works have explored editing via text prompts~\cite{Hertz2022PrompttoPromptIE,Brooks2022InstructPix2PixLT,Cao2023MasaCtrlTM}, spatial guidance~\cite{zhang2023adding}, and other user interactions~\cite{mou2023dragondiffusion}. As the large language models advance, editing paradigms~\cite{Pan2023KosmosGGI, Peng2023Kosmos2GM,Xiao2024OmniGenUI,santos2024leveragingllmsontheflyinstruction} that leverage the model's linguistic reasoning capabilities emerge. While being very powerful, such methods re-generate every pixel in the edited image and hence often struggle with identity preservation (see \Cref{fig:motiovationGenerative}).

\subsection{Reusing LLMs for Graphics Tasks}
We are witnessing a revolution in the domain of (multi-modal) large language models with many successful examples~\cite{touvron2023llama,jiang2023mistral,achiam2023gpt}. These models specialize on various tasks such layout planning~\cite{littlefair2025flairgpt,layoutgpt,open_universe,holodeck},  3D editing~\cite{huang2024blenderalchemy}, and embodied interaction~\cite{qi2024shapellm}. In the context of image editing, ClickDiffusion~\cite{Helbling2024} first generates a new layout given a text prompt from which conditional image generation is performed. Chain-of-thought~(CoT)~\cite{CoT:24} is used at inference time to better leverage the priors of the LLM to create the new layouts at inference time. Fu et al.~\shortcite{mgie24} leverage an MLLM not only to obtain expressive instructions but additional visual guidance to condition a diffusion-based generator, which is fine-tuned (see \Cref{fig:motiovationGenerative}). Ours differs from these methods since we do not cast image editing as a single, closed-box generation but represent it as applying a set of predefined image filters and utilize an MLLM to predict the sequence and parameters of such filters. 

Most recently, several work~\cite{cca2023,wang2024div} explore a framework where an MLLM is used as an agent to plan a sequence of editing operations to be applied given a source image and a target description. The planning is guided by a feedback mechanism in an iterative workflow. In contrast, we fine-tune the MLLM which is then queried at test time directly to generate the editing operations. In a concurrent effort, ComfyGen~\cite{gal2024comfygen} finetunes an LLM to choose a ComfyUI workflow from a given set of flows to accomplish a desired \textit{generative} image editing task. To the best of our knowledge, our method is the first where an MLLM is fine-tuned to reason about a set of procedural image editing operations and their parameters.

\section{Design Considerations}
\label{sec:design}

Our goal is to aesthetically retouch any given image \source using a combination of operations drawn from a library \library of pre-defined procedural filters. We author a library (see supplemental for details) with three types/stages of operations:  
(i)~lighting adjustments (e.g., \texttt{blacks}, \texttt{contrast}, \texttt{exposure}, \texttt{highlights}, \texttt{whites}, \texttt{shadows}); 
(ii)~color and temperatures adjustments (e.g., \texttt{saturation}, \texttt{temperature}, \texttt{tint}); and
(iii)~color-specific adjustments (e.g., \texttt{hue}, \texttt{luminance}, \texttt{saturation}) for eight different colors. 
We execute the adjustments in the three stages, as listed above. %
The ability to fine-tune colors across eight distinct and precise ranges allows us to make meaningful, localized adjustments, providing a practical alternative to using masks for certain editing challenges. We assume that each of the listed functions in the library can be executed by specifying a source image and function parameters (e.g., can be coded a C++, Python, or even neural blocks). 

\paragraph{Designing Visual Puzzles.}
Although MLLMs have rich global priors, they perform poorly when queried directly to generate procedural image retouching operations (see \Cref{fig:motiovationGenerative} and \Cref{sec:evaluation}). Similarly, fine-tuning them to directly generate the parameters of a set of operations given a source image leads to severe over-fitting due to scarcity of training data (see Section~\ref{sec:evaluation}). Instead, \name proposes to make the MLLMs more (image) operation-aware, by designing specific visual puzzles with different goals and training them using suitable datasets. %
Specifically, solving these puzzles helps an MLLM develop the following knowledge:  \\
(a)~{\em what each image operation does}, i.e., relations among a source image, a single (image) operation, and the resultant image; \\
(b)~{\em how much to apply any image operation}, i.e., what is an aesthetic application of an operation on any source/intermediate image; \\
(c)~{\em how far we are from an `optimal' image}, i.e., building an internal model of a desirable retouched image; and finally,  \\
 (d)~{\em how to plan a series of operations to get to an `optimal' image}, i.e., learn how to create an editing plan. \\
With this motivation, we design three puzzles (described in \Cref{sec:method}): Puzzle A helps develop the skill~(a); Puzzle B helps develop skills~(b,c), and Puzzle C helps develop skill~(d).

\textit{Our visual puzzles serve as proxy loss functions for the various image adjustment operations when fine-tuning the MLLM.}

\paragraph{Generating a Reasoning Dataset.}
Once we design the visual puzzles, we use a pretrained MLLM (we use Gemini 2.0 flash) to additionally generate a reasoning solution corresponding to each puzzle. This step allows us to leverage pretrained MLLMs to reason about each edit operation by explaining \textit{why} a particular operation was used and \textit{what} problem it fixes by grounding it on the actual adjustments and basing its reasoning on the visual changes to prevent hallucination. (See \Cref{sec:evaluation} for a comparison with a baseline with Gemini2.0 with our library \library.) We shall use this to build a dataset to fine-tune a MLLM to \textit{acquire this reasoning when the actual adjustments are not supplied}. Next, we describe our method.

\section{Method}
\label{sec:method}

\name is a novel framework that leverages MLLMs for advanced reasoning to facilitate procedural image retouching. Pretrained MLLMs lack the domain knowledge required to comprehend underlying image retouching operations and their associated adjustment values. To address this limitation, we design a set of puzzles specifically targeted at bridging these knowledge gaps. We discover that by solving these puzzles, the MLLM can become an agent with expert-level domain knowledge, capable of retouching images effectively. In the following, we first introduce the visual puzzles we design to fine-tune the MLLM, and then discuss how we utilize the fine-tuned model at inference time. Finally, we provide details of the procedural image filters we use and how we execute them in practice.

\subsection{Puzzle A: Gaining Understanding of Individual Operations} \label{sec:puzzle1}

As the first step, the MLLM must visually understand the impact of each individual operation on an image and how this impact varies across different adjustment levels. To do this, we randomly sample an operation \operation $\in$ \library and an associated adjustment value \adjustvalue from a predefined library of image retouching operations. This operation is then applied to a source image \source to produce an edited image \edited. The \source and \edited images are stitched together and presented to an MLLM to identify the operation and adjustment value, i.e., given the image pairs the MLLM should predict the operation and its corresponding change amount.

\begin{figure}[t!]
    \centering
    \includegraphics[width=\columnwidth]{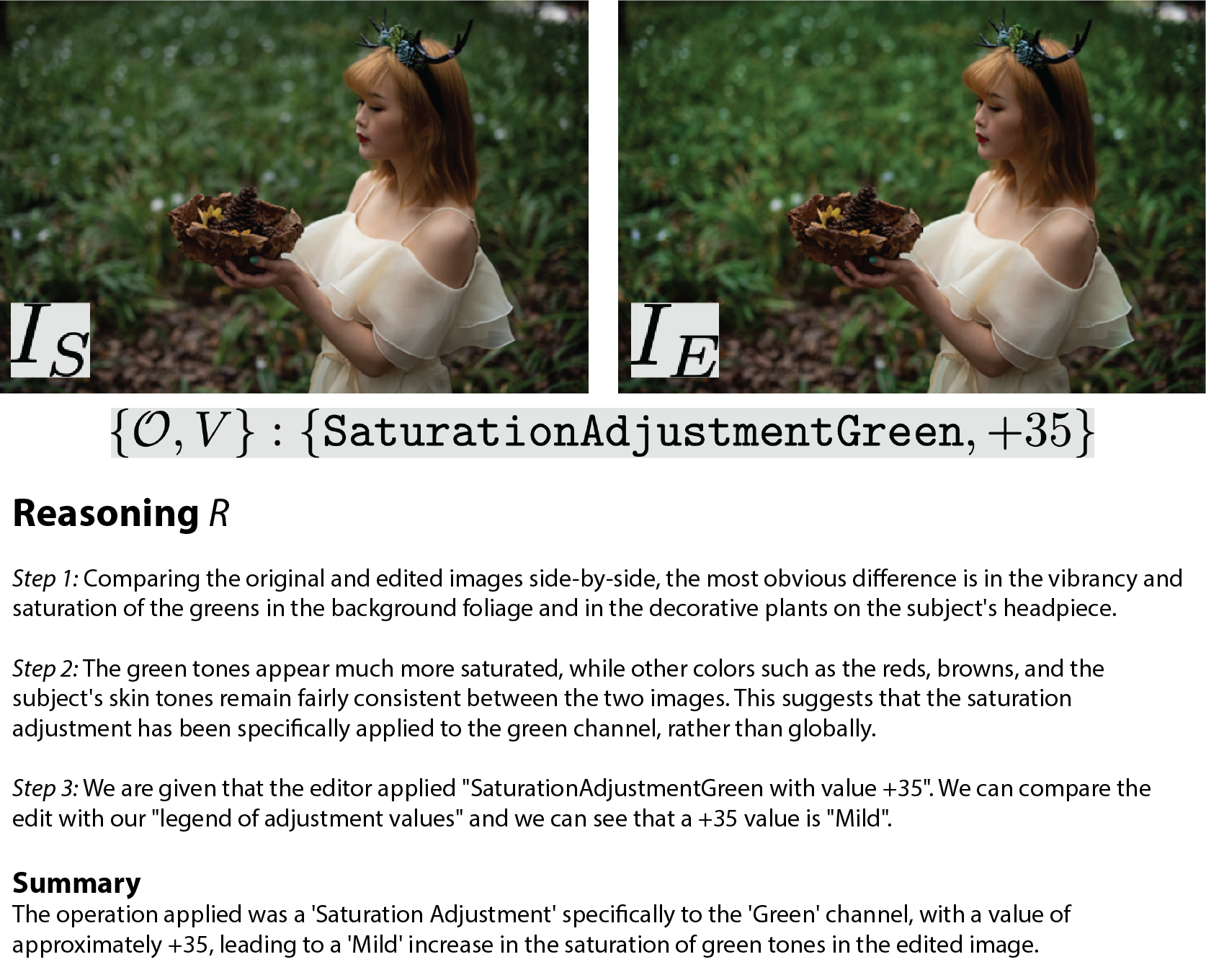}
    \caption{\textbf{Puzzle A}. This puzzle intends to teach what any single operation \operation $\in$ \library, along with its value \adjustvalue, does to a source image \source to produce an edited image \edited. The visual puzzle being, given an ordered pair $(I_S, I_E)$, one has to predict $(\mathcal{O},V)$. Using $(I_S, I_E)$, we also generate corresponding reasoning \reason. 
    }
    \label{fig:puzzleOne}
\end{figure}

We discover that directly querying a pretrained MLLM to identify the operation and adjustment value produces poor results (presented as `Gemini 2.0+CoT+library' option in \Cref{sec:evaluation}). However, when provided with the actual operation and adjustment value and grounded on the observed visual changes, the MLLM generates detailed and convincing reasoning \reason that accurately explains in rich textual descriptions, how the effect of ($\mathcal{O}$,\adjustvalue) on \source results in \edited. Thus, given pairs $(I_S,I_E)$, we get associated \reason as well as (single) edit ($\mathcal{O}$,\adjustvalue), see  \Cref{fig:puzzleOne}. Grounding the MLLM with the true adjustment value ensures its reasoning aligns with the specific operation \operation, avoiding unrelated or hypothetical explanations.

We use the extracted reasoning during the training phase to teach the MLLM in a supervised fashion to identify \operation and regress \adjustvalue. (Note that we synthetically generate this training data as we have access to the image processing filters in \library.) Instead of directly inferring the operation and adjustment value, the MLLM needs to explain the reasoning by eliciting visual differences in the image caused by \operation in tandem with the degree of adjustment. By elaborating on the visual changes in the edited image, the \textit{MLLM encodes these visual nuances within its text representation}, enabling it to effectively learn the effects of various operations in library \library.
We quantized (normalized) parameter values (see also \cite{wang2024llamameshunifying3dmesh}) as they are easier to tokenize and use with MLLMs.

\begin{figure}[b!]
    \centering
\includegraphics[width=\columnwidth]{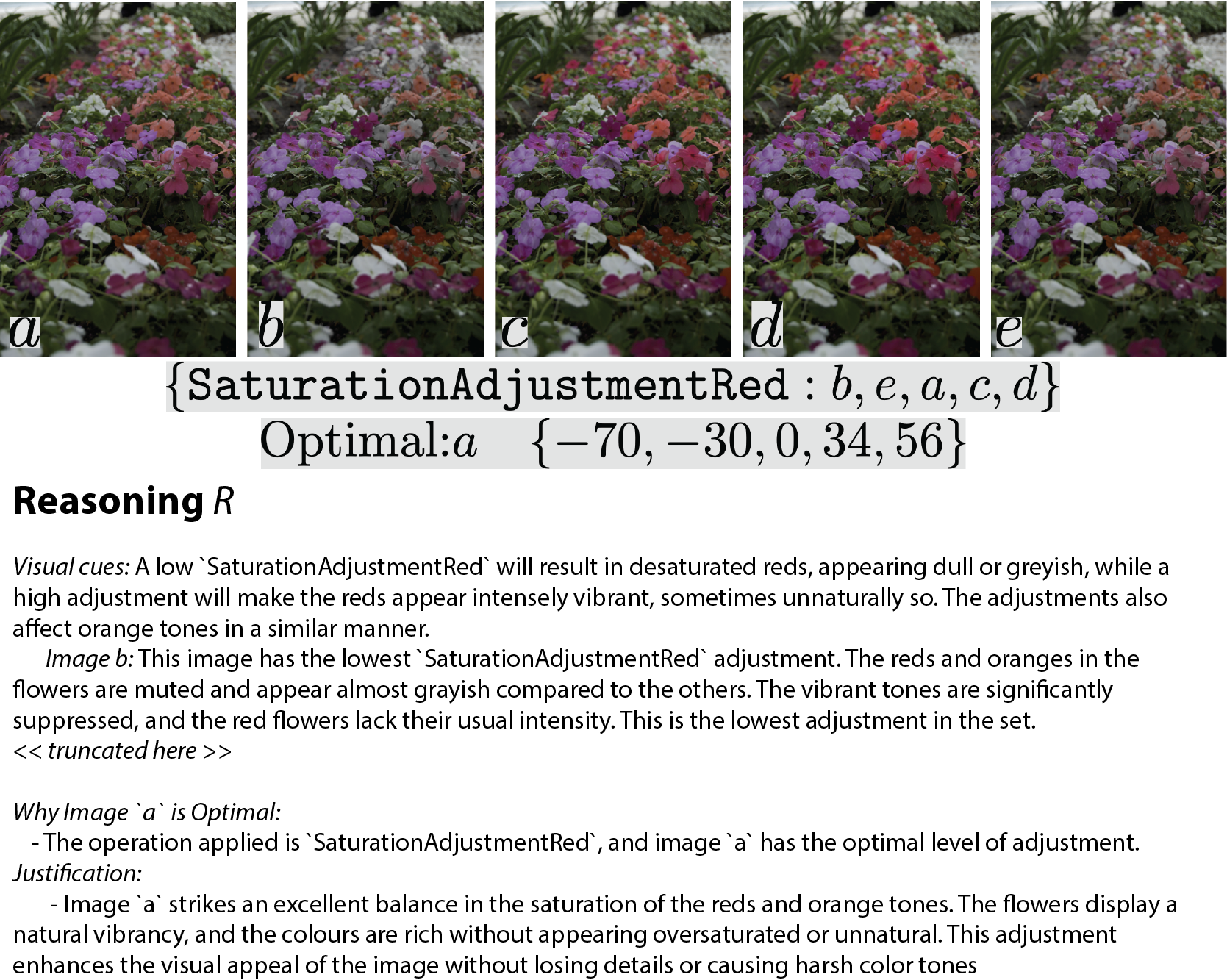}
    \caption{\textbf{Puzzle B}. This puzzle intends to teach about image aesthetics under any single operation \operation $\in$ \library. The visual puzzle being, given a set of randomly ordered images $(I_E, I_{V_1}, I_{V_2}, I_{V_3}, I_{V_4})$, generated from an expert-edited final image \expert by applying operation \operation with perturbed values $\{V_i\}$, one has to 
    order the set of images, based on low to high values $V_i$, as well as identify the optimal image \expert along with the perturbations values to go back to \expert from each image. Note that we implicitly assume that any perturbation of an expert-edited image results in a worse image. Using the image set and the operation, we also generate corresponding reasoning \reason. }
    \label{fig:puzzleTwo}
\end{figure}

\subsection{Puzzle B: Understanding Image Aesthetics} \label{sec:puzzle2}
Understanding image aesthetics is essential for defining how an enhanced version of an image should appear. An MLLM must acquire the capability to visually identify the optimal appearance of an image when an operation is adjusted to its ideal parameter value. To achieve this, we design a second puzzle involving four random variations ($I_V$) of adjustments for a sampled operation, \operation, applied to an image edited by an expert, \expert. Note that we assume that any sufficiently large adjustment made to \expert degrades the image and results in a suboptimal edit.

We construct the puzzles by stitching \expert alongside the four adjusted images in a random order. The MLLM is first tasked with ordering all five images from the lowest to highest adjustment values. The adjustment range for the operations is defined on a  perceptually linear scale, ranging from $[-100,+100]$. Once ordered, the MLLM must then identify the image with the optimal level of \operation (i.e., identify \expert) and justify its reasoning. Additionally, it must determine the adjustment level required to transform a randomly selected image from $I_V$ into the optimal image, \expert. See \Cref{fig:puzzleTwo}. Note that this process implicitly assumes that the operation is invertible. 

Similar to \Cref{sec:puzzle1}, instead of providing a direct one-sentence answer to the puzzle, we query a pretrained model with the correct answers and task it with generating reasoning grounded in the observed visual changes. By training the MLLM to solve this puzzle and detail its reasoning, it acquires an intrinsic ability to recognize the visual characteristics of an optimally adjusted image and estimate the adjustment value required to transform any source image to the optimal one. This aesthetic understanding is critical when planning edits that involve multiple operations and adjustments.

\subsection{Puzzle C: Generating a Plan for Image Retouching} \label{sec:puzzle3}
MLLMs have demonstrated significant success in solving complex tasks, such as mathematical proofs, by decomposing them into manageable steps~\cite{CoT:24}. However, the abstract and subjective nature of image editing presents a large state space, making it challenging for an MLLM to directly predict multiple operations along with their corresponding adjustment values. While existing MLLMs can suggest basic adjustments, such as modifying exposure or saturation, they struggle to produce comprehensive editing plans with precise adjustments.

\begin{figure}[b!]
    \centering
\includegraphics[width=\columnwidth]{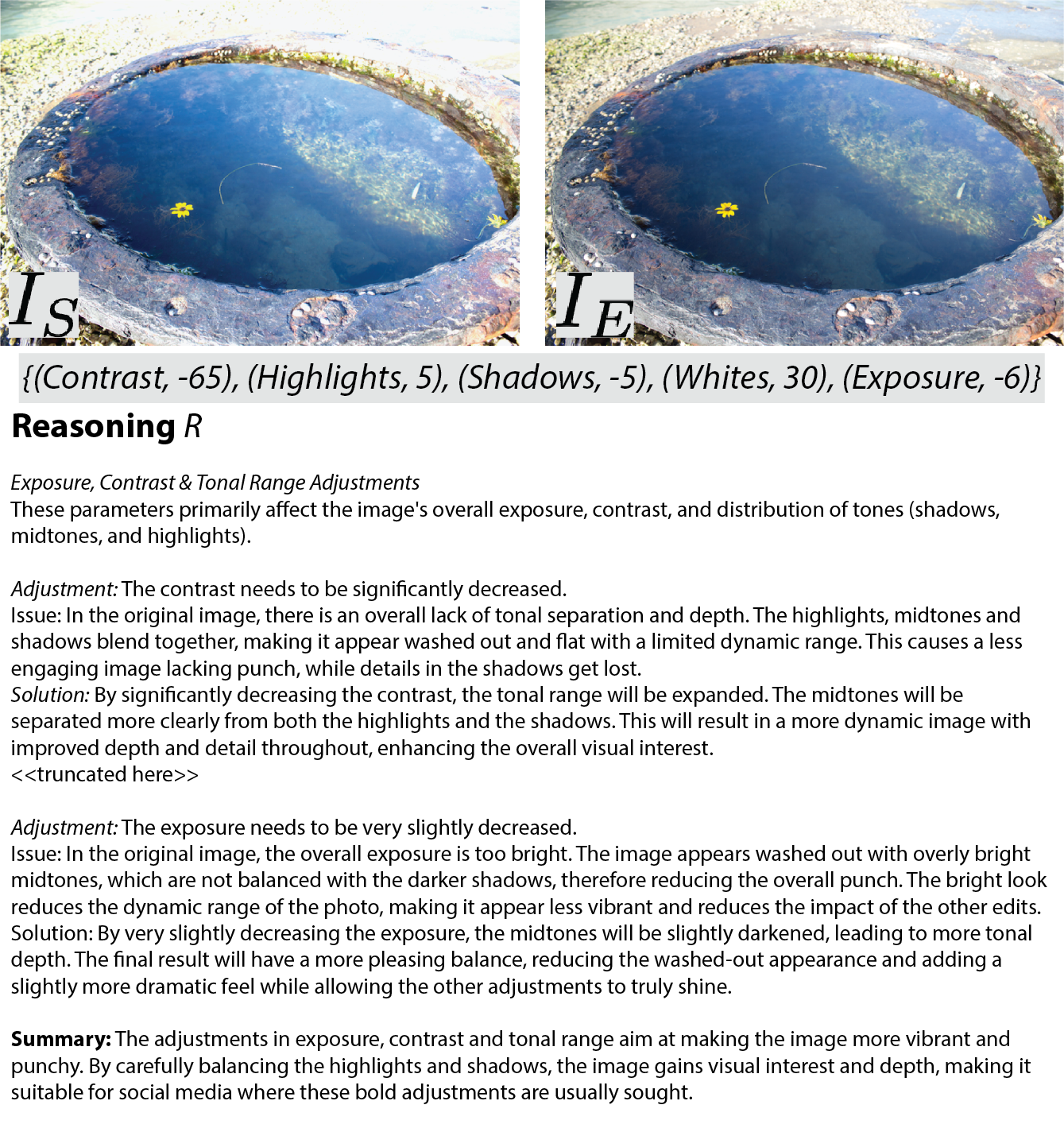} 
\caption{\textbf{Puzzle C}. This puzzle intends to teach how to generate a retouching plan. The visual puzzle being, 
given an ordered pair $(I_S, I_E)$, one has to come up with a retouching plan $\{(\mathcal{O}_i, V_i)\}$ listing the operations, from \library, along with the associated parameter values. Using the image retouching sequence and the operations, we also generate corresponding reasoning \reason, in the form of <\texttt{Adjustment}, \texttt{Issue}, \texttt{Solution}> for each operation. 
    }
    \label{fig:puzzleThree}
\end{figure}

To bridge this gap, we design a third puzzle, where we aim to enable MLLMs to generate expert-level retouching plans, including suitable operations and adjustment values, to enhance a source image \source into its expertly retouched counterpart \expert. To achieve this, we modify expert-edited images to create poorer-quality variants, which serve as the \source images requiring enhancement. Note that, once again, we only make use of expert-edited images \expert, while we use synthetic edit plans by procedurally perturbing them. However, unlike previous puzzles which focused on learning operations individually, we alter multiple parameters within specific categories:
(i)~Lighting adjustments, 
(ii)~Color and temperature adjustments, and 
(iii)~Color-specific adjustments.

For the planning stage, our design choice is motivated by the following considerations:
(i)~\textit{Inversion feasibility:} By modifying a limited set of parameters within a category, the operations remain invertible, enabling the reconstructed image to closely match \expert.
(ii)~\textit{Reduced complexity:} Generating a comprehensive plan involving numerous operations simultaneously is inherently challenging. Dividing the process into sequential stages—starting with lighting, followed by color-temperature adjustments, and finally color-specific fine-tuning—simplifies the task and aligns with expert workflows.
(iii)~\textit{Reasoning clarity:} Finally, similar to the first two puzzles, we want to generate reasoning behind the edits by querying a pretrained MLLM to analyze visual changes corresponding to specific adjustments. When many operations are applied simultaneously, it becomes challenging to disentangle which operation is responsible for a given visual change. Hence, we split our task into stages. 

We synthetically generate a dataset of \source images derived from expert edits \expert. For each \source-\expert pair, we task the MLLM with generating reasoning for every adjustment. This involves identifying the adjustment to be applied, noting the degree of change, and linking it to the corresponding visual issue and solution. Specifically, we structure the reasoning into a triplet <\texttt{Adjustment}, \texttt{Issue}, \texttt{Solution}> for each operation, as described next (see \Cref{fig:puzzleThree}).

\noindent
\texttt{Adjustment:} The operation and its degree of adjustment.

\noindent
\texttt{Issue:} The visual problem addressed by the adjustment by referencing specific elements in the image.

\noindent
\texttt{Solution:} The visual improvement achieved through the adjustment.

We reformatted the detailed triplets in the style of instructions to generate a plan \plan. \textit{Note, during training the MLLM only has access to \source and it must generate a plan, \plan, that will lead to \expert}. 
When fine-tuned on this dataset, the MLLM is capable of generating comprehensive plans during inference, without hallucination. Additionally, the MLLM uses chain-of-thought~(CoT) reasoning to regress from high-level reasoning to precise parameter values, referencing a predefined legend that maps adjustments to numerical ranges. This structured approach ensures that the MLLM produces meaningful insights and solutions aligned with expert edits, instead of independently guessing adjustment plans.

An important aspect of this puzzle involves teaching the MLLM to recognize when no further adjustments are needed for a particular stage. This prevents unnecessary edits that could degrade a well-adjusted image. To train this skill, we introduce an additional challenge: the MLLM when queried to generate an editing plan for a given stage, it must instead justify why no further edits are required. As in earlier puzzles, synthetic reasoning is generated to train the MLLM by assuming that any further modification to an expert-edited image \expert for a specific category of operations would result in a degradation of quality.

\begin{figure*}
    \centering
    \includegraphics[width=\textwidth]{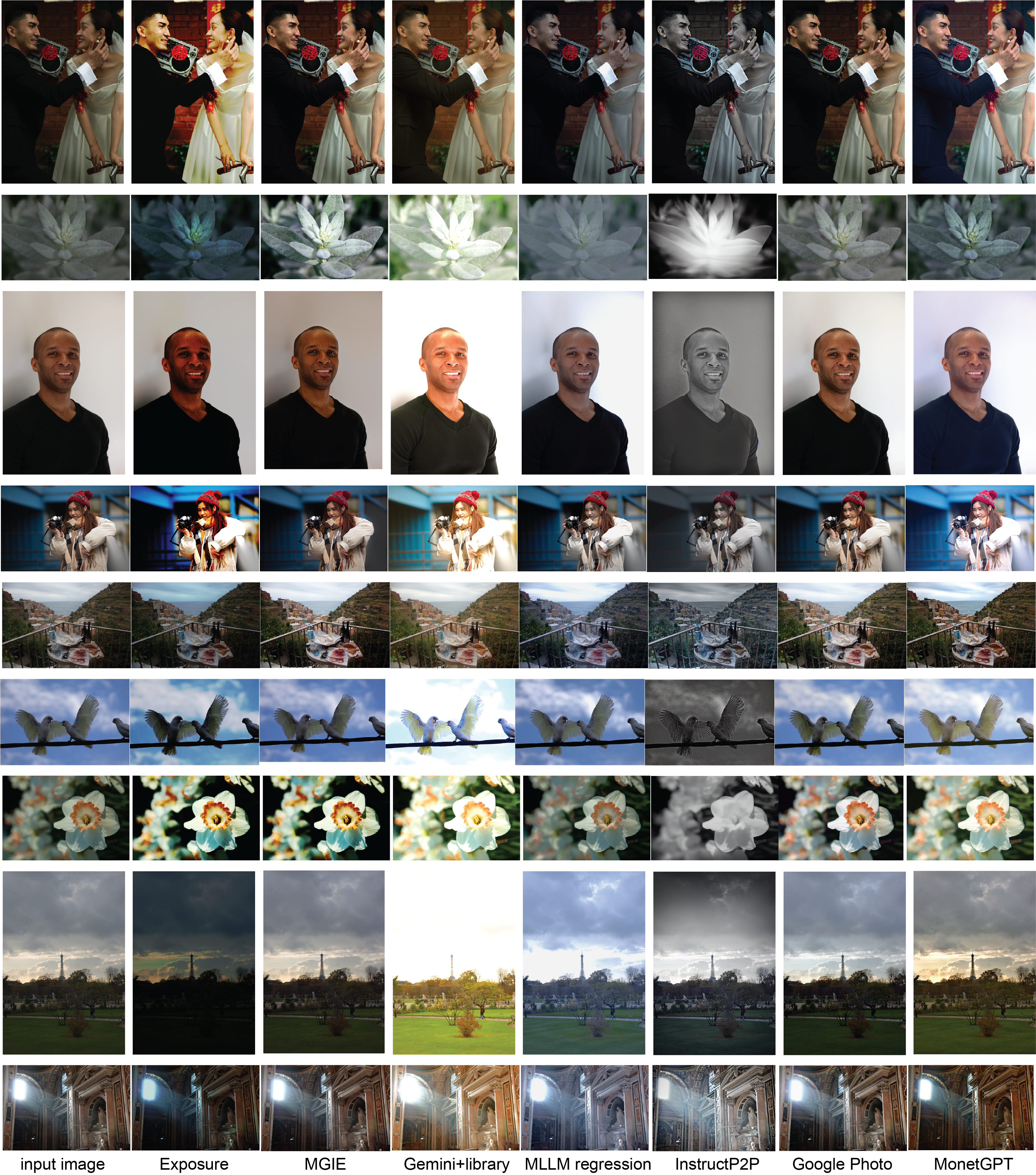}
    \caption{Each row depicts an input image and a retouched version produced by each of the baselines. While generative baselines (MGIE) struggle to preserve the identity (last row), Exposure or Gemini sometimes generate too bright or dark results (third row). Direct regression with an MLLM fails to make sufficient enhancements. Our method is capable of providing balanced and aesthetic enhancements.}
    \label{fig:resultsGallery}
\end{figure*}

\subsection{Inference: Reasoning as a Pathway for Regression}
Once the MLLM is trained on the three puzzles, we use it to generate editing plans by leveraging reasoning. Reasoning bridges the gap between adjustment values and the underlying intent behind each editing operation. This effectively induces a pathway for the fine-tuned MLLM \llm to regress precise adjustment values regress from high-level reasoning. Given \source, our goal is to predict a set of adjustment operations and corresponding values, \adjust. We condition an MLLM \llm on \source to first generate an editing plan \plan as, 
\begin{equation}
    \mathcal{M}(\cdot| I_S) := \mathcal{P}.
\end{equation}
We then drive \llm to generate the final adjustment values \adjust based on \plan,
\begin{equation}
    \mathcal{M}(\cdot| \mathcal{P}, I_{{S}}) := A.
\end{equation}

As noted in \Cref{sec:puzzle3}, we divide our plan generation into three separate stages. Hence, we apply our procedural pipeline on the source image to get an edited image, which we feed back to infer the next stage of operations, as dictated by the inferred plan. See \Cref{fig:teaser} for an example.

\subsection{Authoring a Library with a Reduced Parameter Space}  
Existing image enhancement software with Python bindings, such as GIMP, are overly complicated to code for. For instance, GIMP's procedural database (PDB) often requires multiple function calls and the specification of numerous parameter values to execute a single adjustment operation, unnecessarily expanding the parameter space. Other Python-based image enhancement libraries, such as OpenCV and Pillow, provide a more straightforward interface but offer a very limited range of operations.  

To address these limitations and making use of open source options, we developed a Python library of image adjustment operations that simplifies the process of defining and executing adjustments. The library uses modular functions, where each adjustment operation is controlled by a single master parameter. Sub-parameters are either fixed or dynamically derived from the master parameter, significantly reducing the complexity of the parameter space. The library provides an approximately comparable subset of tools to those available in platforms like Google Photos and Lightroom. 

More importantly, the fine-tuned MLLM possesses a deep visual understanding of the operations within the library, enabling it to generate detailed plans with adjustment values that accurately reflect how each modification impacts the image. 

Leveraging this understanding, the fine-tuned MLLM acts as an agent that generates editing plans by generating operations and corresponding adjustment values as output in a structured JSON format, eliminating the need to write code and instead solely focus on capturing the visual impact of each adjustment. Our library directly processes this JSON file to apply the adjustments seamlessly. Additionally, we design the parameter values to follow a mostly perceptually linear scale, ranging from [-100,+100] to ensure a consistent control over the adjustments.
Our operations are non-destructive by construction and can operate on high-resolution 16-bit images compared to other generative solutions.

\begin{figure*}[t!]
    \centering 
\includegraphics[width=\textwidth]{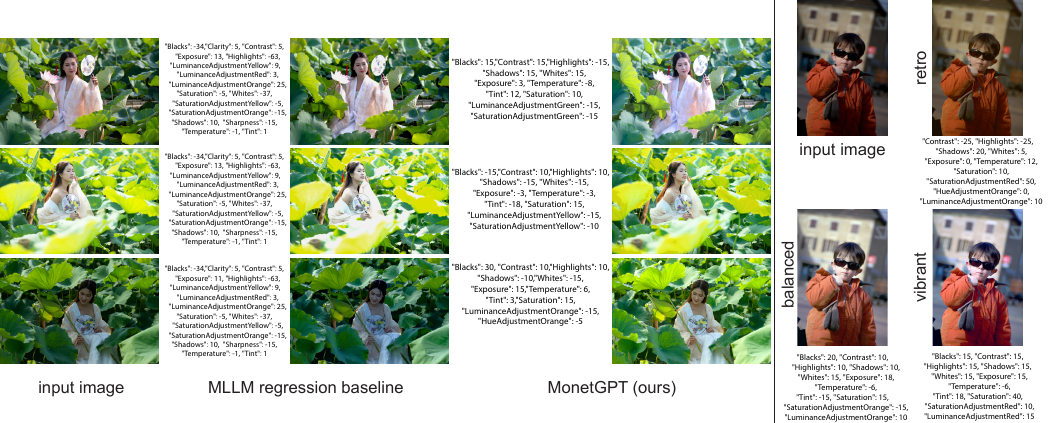}
    \caption{(Left)~\name responds to subtle changes in the input image, producing different edit plans. In contrast, our regression baseline largely misses the subtle variations in the input and proposes almost identical retouching plans. 
    (Right)~\name can produce different edit plans based on style tags, here we show \texttt{retro}, \texttt{balanced}, and \texttt{vibrant} producing different edit plans, resulting different retouched results. 
    }
    \label{fig:varietyOfEdits}
\end{figure*}

\section{Evaluation}
\label{sec:evaluation}
\textbf{Implementation Details.}
We fine-tune Qwen-VL-7B-Instruct \cite{Qwen2VL}, a 7-billion-parameter MLLM (\llm), using DoRA adapters \cite{dora}. We fine-tune using the \texttt{llamafactory} ~\cite{zheng2024llamafactory} framework. We configure DoRAZ\cite{dora} with a dropout rate of $0.2$, an adapter rank of $256$, and an alpha rank of $512$. A learning rate of $1\text{e}{-4}$ is employed with cosine scheduling, and the model is fine-tuned for a single epoch, which takes approximately 8 hours on an H100 GPU; direct regression baseline takes 2.5 hours. The training dataset comprises synthetically generated puzzles paired with corresponding reasoning, generated using Gemini 2.0 Flash Experimental~\cite{gemini}. We created the puzzles by sampling expert edited images from PPR10K~\cite{ppr10k}. As detailed in \Cref{sec:method}, we apply random adjustments to each image, yielding a synthetic puzzle dataset with approximately 7k samples for puzzle A, 5k for puzzle B, and 13k for puzzle C. Our library includes a total of 33 operations, implemented in Python either from scratch or by extending existing libraries such as OpenCV and Pillow (see supplemental). To ensure the accuracy of our library, we verify that applying the expert adjustment values from the PPR10K dataset to the source images produces results that closely match the desired targets. 

\textbf{Inference time}. 
For inference, we used an RTX 4090. For each retouching, ours takes 25~sec for full staged pipeline execution, while direct regression takes 10~sec, and Exposure takes about 2~sec.

\textbf{Dataset}. We evaluate our method on a  variety of images curated from the PPR10K~\cite{ppr10k} and Adobe5k~\cite{Bychkovsky2011} datasets,  which provide access to source images as well as expert edits. Recall that our data generation does not make use of the pairing information. For testing, we select images for which the expert versions have not been seen during our training.

\textbf{Baselines.} We perform extensive comparisons with the following methods. (i) \textit{Exposure}~\cite{hu2018exposure}, which is an RL-based framework to suggest a sequence of operations and their parameters to enhance an image; (ii) \textit{Unpaired Image Enhancement}~\cite{kosugi2020unpaired}, similar to Exposure, with an editing interface; (iii) RSFNet~\cite{rsfnet}, which generates pixel-level attention maps using region-specific filters but it requires paired retouched images for training that may be hard to obtain; (iv) \textit{MGIE}~\cite{mgie24} utilizes an MLLM to derive expressive instructions and additional guidance to enable instruction guided image editing. In our experiments, we use the fixed instruction \textit{`Enhance the image like a professional image editing expert using Lightroom'}; (v)~\textit{Gemini+library}, where we use Gemini 2.0  at inference time with chain-of-thought (CoT) reasoning along with our library \library, but with no additional training or fine-tuning. In particular, we interact with Gemini in three stages. Given a source image, we first prompt it to write a detailed plan of the adjustments to be made without providing any operation names or parameters. This effectively enforces the MLLM first to reason about what aspects of the image require enhancement. We then provide three categories of operations, i.e., lighting adjustments, color and temperature adjustments, and color-specific operations, similar to ours. We prompt the MLLM to choose which operations need to be applied along with a reasoning.  
Finally, we ask the MLLM to provide the parameters of the selected operations,  which we then translate and execute using our library. We refer to the supplementary material for more details about each step of this prompting; 
(vi)~\textit{InstructP2P}~\cite{brooks2022instructpix2pix}; and finally
(vii)~\textit{Google Photo}, where we also use the autoenhance feature available as part of Google Photos\footnote{https://photos.google.com/} as a black box commercial alternative. 

Finally, to ablate our method, we also present \textit{regression} that refer to an MLLM guided variant where we directly fine-tune the MLLM (same setup as ours) to directly regress the parameters of a set of image editing operations. We perform this fine-tuning using the PPR10K dataset where we have access to source and edited image pairs along with corresponding adjustment operations and values. Following prior works such as Exposure~\cite{hu2018exposure}, we train it to emulate a single expert's adjustments to minimize ambiguity.

\setlength{\tabcolsep}{5pt}
\begin{table}[htbp] %
  \centering
  \caption{Evaluation for generalization on the Adobe5k dataset. All methods are trained on the PPR10k dataset except MGIE (requires very large data) and GooglePhotos (closed source and proprietary).}
  \label{tab:adobe5k}
  \hspace{-0.1in}\resizebox{0.99\linewidth}{!}{
  \begin{tabular}{r c c c c }
    \toprule %
    Method             & SSIM ↑      & LPIPS ↓      & PSNR ↑      & Histogram ↑ \\ %
    \midrule %
    Exposure~\cite{hu2018exposure}          & 0.63        & 0.14         & 15.12       & 47.21          \\ %
    Unpaired~[Kosugi et al.~\citeyear{kosugi2020unpaired}]           & 0.83       & 0.12        & 21.73      & \underline{83.98}       \\   %
    RSFNet~\cite{rsfnet}           & \underline{0.88}       & 0.08          & 21.85      & 80.26       \\   %
    InstructP2P~\cite{brooks2022instructpix2pix}               & 0.61        & 0.22         & 16.99       & 73.90          \\ %
    MGIE~\cite{mgie24}               & 0.74        & 0.08         & 22.94       & 79.95          \\ %
    GeminiCoT ~\cite{gemini} + our library        & 0.80        & 0.14         & 17.83       & 63.71           \\ %
    GooglePhotos       & \textbf{0.90} & \textbf{0.06}  & \textbf{25.86} & \textbf{86.47} \\ %
    MLLM Regression    & 0.84        & 0.10         & 20.89       & \textit{82.05} \\ %
    Ours               & \textbf{0.90} & \underline{0.07}  & \underline{23.75} & 79.50       \\  %
    \bottomrule %
  \end{tabular}}
\end{table}

\begin{figure*}[t!]
    \centering
    \includegraphics[width=0.82\textwidth]{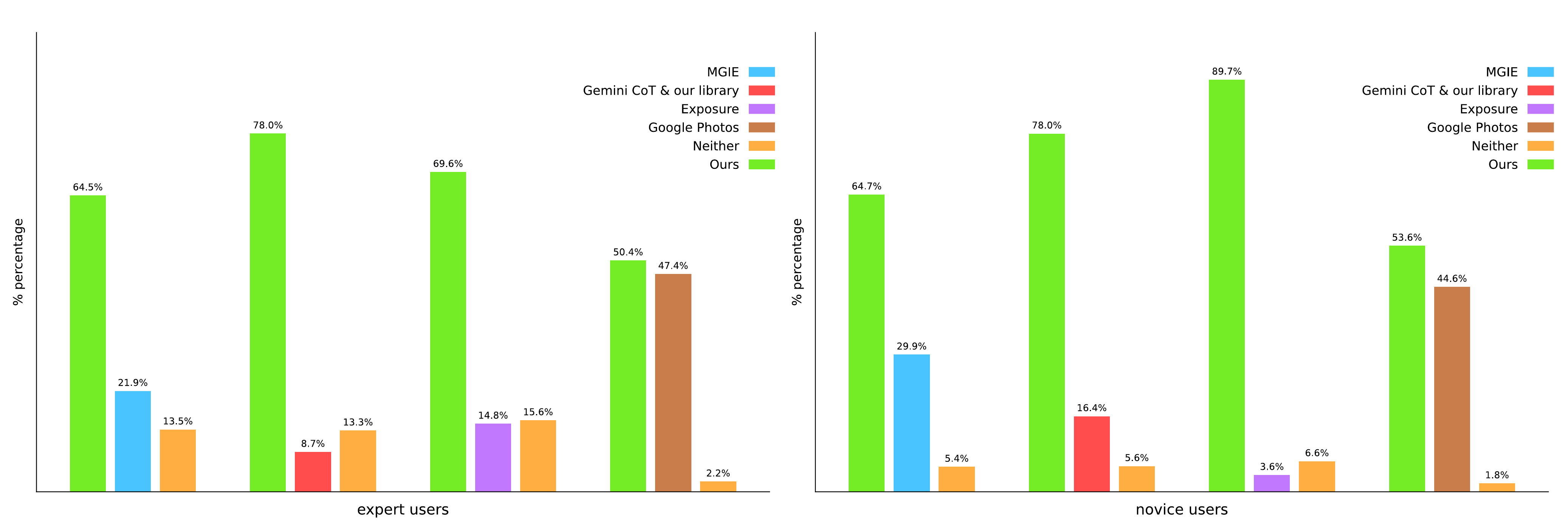}
    \caption{User preference study comparing our method against baselines Exposure~\cite{hu2018exposure} (white-box system) and MGIE~\cite{mgie24} (instruction-guided MLLM enhancer). We presented participants with a source image alongside a pair of edited images, where our result was randomly paired with one of the baselines. Responses were collected from users experienced in photo retouching (\textit{expert} users) and those with varying familiarity (\textit{novice} users). Participants could also select \texttt{neither} if both edits failed to improve upon the original image. Both expert and novice groups demonstrated a preference for our results, as shown.}
    \label{fig:results_user_study}
\end{figure*}

\textbf{Quantitative Comparison.} We train MonetGPT and baselines on a single expert (expert A) from PPR10k~\cite{ppr10k} and evaluate on 400 images randomly sampled from Adobe5k~\cite{Bychkovsky2011} to test for \textit{generalization}. We do not re-train MGIE due to its significantly larger data requirements (1M+), nor Google Photos, which is closed source. We also note that training solely on PPR10k makes generalization more challenging due to its limited image variety (mostly portraits). For evaluation, we compute several standard metrics: PSNR measures pixel-wise fidelity, while SSIM and LPIPS~\cite{lpips} serve as perceptual quality indicators. We additionally compute histogram intersections based on Hu et al.~\shortcite{hu2018exposure} to assess how well the predicted image distributions match expert edits for contrast, luminance, and color saturation (we show the mean of the three histograms in \Cref{tab:adobe5k}). The Adobe5k dataset provides edits by five different experts. For each sample and metric, we take the highest score achieved against any of these five experts (except for histogram intersections, which consider all experts). Given the subjective nature of image retouching, matching any expert edit can be considered a desirable outcome for an edited sample. Results in \Cref{tab:adobe5k} show that our method outperforms all open-source baselines on three of the four metrics and achieves performance comparable to the closed-source Google Photos.

\textbf{Qualitative Comparison.} Traditional image comparison metrics often fail to capture the enhancement quality given the subjective nature of image retouching. Therefore, we conduct user studies and expert judgment to evaluate our results. In particular, we first conduct a user study where we select 15 source images and generate enhanced versions using Exposure, Gemini-CoT, and our method. We present each participant the source image as well as a pair of enhanced results and ask them to choose the option that has better aesthetic quality and visual enhancement. They are also provided with the option of choosing neither of the options if they think the source image looks aesthetically more plausible. Each pair is constructed from our result as well as one of the baselines and presented in random order resulting in a total of 200 questions. We collect answers from 15 novice users with varying skills in photo retouching as well as 10 photography experts. 

\begin{figure}[b!]
    \centering
\includegraphics[width=0.9\columnwidth]{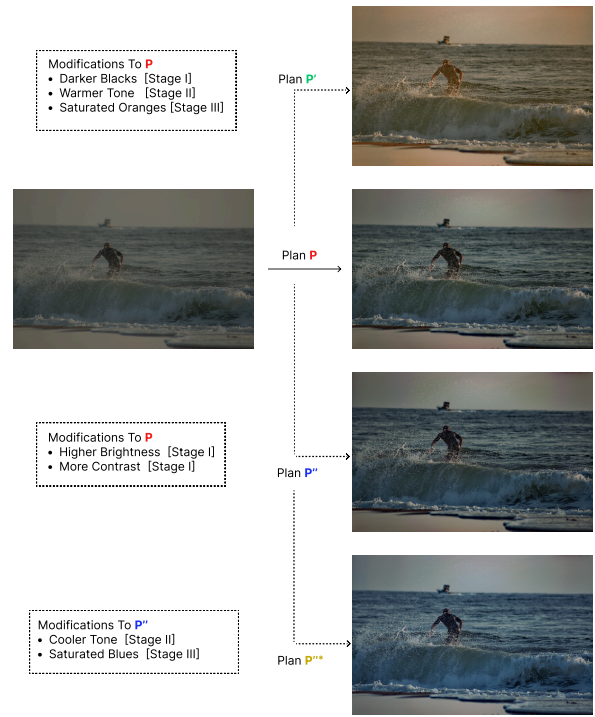} 
\caption{\textbf{Autoregressive editing}. The autoregressive nature of MLLMs, combined with our staged editing pipeline, allows the user to edit the plan (\plan) at any stage. The refined plan is used to determine subsequent parameter values. Moreover, the edited plan $\mathcal{P^{'}}$ enables the MLLM to generate plans for subsequent stages that are consistent with the edits. The bottom image ($\mathcal{P^{''*}}$) shows a result following further modifications to the second and third stages, after the first stage was modified to generate $\mathcal{P^{''}}$.}
    \label{fig:autoregressive}
\end{figure}

\begin{figure*}[t!]
    \centering 
\includegraphics[width=\textwidth]{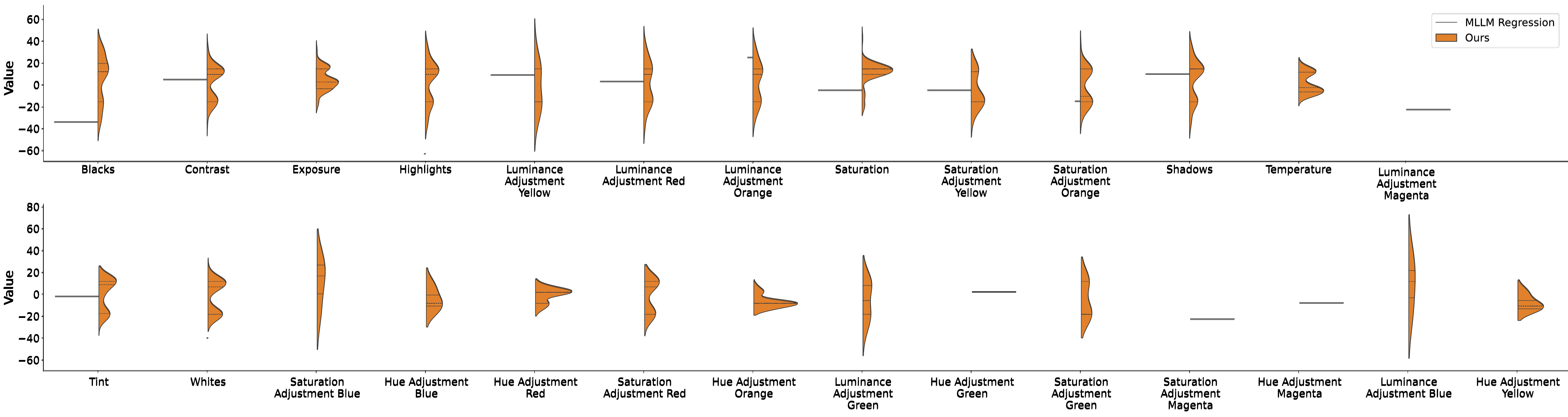}
    \caption{For each operation in our library, we show a violin plot of the adjustment values proposed by \name and the baseline which directly regresses the values on 100 images from the PPR10k dataset. While the baseline overfits and predicts the same values, \name utilizes the full range of values.}
    \label{fig:violinPlot}
\end{figure*}

\textbf{Discussion of results.}
We provide a sample set of results in Figure~\ref{fig:resultsGallery}, and refer to the supplementary material for more examples including detailed explanations for our edits. IP2P and MGIE, being generative methods, are limited in terms of the resolution of images it can generate and often struggle to preserve the content of the source image. Exposure and Gemini CoT+library baselines frequently result in over-exposed, high contrast, too bright or too dark images (e.g., rows 2 and 8). The \textit{auto enhance} option in Google Photos is a strong baseline, which also suggests parameters of various image editing operations, most likely based on machine learning methods. Our MLLM regression baseline fails to learn any meaningful signal from the paired dataset of adjustment settings (note: ours is unpaired), which further confirms the need for using reasoning as a pathway.

\textbf{Perceptual User Study}.  
Our visual observations are confirmed by our user study on 50 images from-- Adobe5k and Reddit. As shown in Figure~\ref{fig:results_user_study}, our method is preferred against all baselines, both by 15 novice users as well as 10 experts. Beyond completing the user study, we also collected verbal feedback from the experts regarding what worked well and areas for potential improvement in the edits. Overall, the experts expressed a strong preference for our approach, while providing constructive suggestions for minor refinements, such as \textit{`making the skin tones slightly more saturated'} and \textit{`curve adjustments to better emphasize focal points'}.

\textbf{Effect of Image Operation-Awareness}.  
To highlight the impact of puzzle-solving on image operation-awareness, we test \name on the same scene with variations in lighting (primarily), as shown in \Cref{fig:varietyOfEdits}-left. Unlike our MLLM regression baseline, \name generates significantly different retouching plans, tailored to the lighting conditions of each image. In this example,  \name adjusts its outputs, producing distinct adjustments for inputs with balanced lighting, overly bright conditions, or underexposed.

\textbf{Personalized retouching}. Image retouching is subjective and there is no single `optimal' solution, as stylistic preferences vary widely. Our framework is trained primarily to emulate a particular expert's style, which can introduce subjectivity of a single expert when considering an edit as optimal. However, our framework's inherent flexibility—enabled by combining MLLMs with a procedural design—allows it to generalize effectively to diverse stylistic requests specified by users. Users can guide the retouching process via natural-language (e.g., requesting increased vibrancy or softer tones), making the model adaptable to individual preferences. As shown in \Cref{fig:varietyOfEdits}-right: we show three different styles, enabled by providing the following additional \textit{tags} to \name as prompts: `nostalgic retro vibe', `balanced', and `vibrant and punchy colors', and applied to the same input. We query the LLM to characterize the features of a particular \textit{style tag} and then add them to our template prompts. The autoregressive nature of MLLMs further allows us the user to edit the plan at any stage and generate subsequent parameter values and plans for the next stages in tandem with the changes made by the user as presented in \Cref{fig:autoregressive}.

\section{Conclusion}

We demonstrated that MLLMs can learn procedural image retouching operators through training on specially designed visual puzzles. Once trained, the MLLM can critique photographs, propose fixes, and suggest sequences of retouching operations with corresponding parameters. These suggestions can then be translated into executable calls using a function library. We evaluated our approach, \name, on benchmark datasets, demonstrating advantages over various alternatives. Notably, our method requires no inference-time optimization (e.g., iterative feedback), is compatible with existing MLLMs, and is explainable by design (with detailed reasoning).

\textit{Limitations and Future work.}

(i)~Currently, \name supports a limited set of global operations, excluding crops or regional edits. Supporting object-specific operations could involve using semantic masking networks to pre-segment images. However, obtaining sufficient artist-edited images linked with regional masks remains a challenge. 

(ii)~We trained \name on a dataset of 8k expert-edited images. Consequently, artist-specific aesthetic priors or biases are likely reflected in our model. Training on larger, more diverse datasets could mitigate bias, enable learning priors over editing parameters, and potentially facilitate developing aesthetic scoring models.

(iii)~Image retouching is subjective, lacking a single optimal solution. Our model can make errors, sometimes resulting in artifacts like saturated regions. We expect improved training data and better modeling of perturbation distributions in synthetic augmentation to partially address this. A human-in-the-loop system could further enhance user satisfaction.

(iv)~Our work focused on procedural operations and did not include generative filters. Future work could explore casting specific generative edits as neurosymbolic modules, allowing \name to incorporate them. However, this might compromise the procedural interpretability that is a key advantage of our current system.

\begin{acks}
We thank Rishabh Kabra, Ruchira Ray, Tobias Ritschel, Chen Liu, Sylvain Paris, and Morten Hannemose for their comments and suggestions. Niloy was supported by gifts from Adobe and UCL AI Centre.
\end{acks}
\bibliographystyle{ACM-Reference-Format}
\bibliography{main}

\end{document}